\documentclass[fleqn,10pt]{wlscirep}
\pdfoutput=1
\usepackage[utf8]{inputenc}
\usepackage[varg]{txfonts}
\usepackage{graphicx}
\usepackage{color}
\usepackage{setspace}
\newcommand{\chg}[1]{\textcolor{black}{#1}}

\newcommand{\tabrule}{\rule[-0.2em]{0em}{1.2em}}

\title{Swift thermal steering of  domain walls in  ferromagnetic MnBi stripes}

\author[1]{Alexander Sukhov}
\author[1]{Levan Chotorlishvili}
\author[2]{Arthur Ernst}
\author[2]{Xabier Zubizarreta}
\author[2]{Sergey Ostanin}
\author[1,2]{Ingrid Mertig}
\author[2]{Eberhard K. U. Gross}
\author[1,*]{Jamal Berakdar}
\affil[1]{Institut f\"ur Physik, Martin-Luther-Universit\"at, Halle-Wittenberg, D-06099 Halle/Saale, Germany}
\affil[2]{Max Planck Institute of Microstructure Physics, D-06120 Halle/Saale, Germany}

\affil[*]{Jamal.Berakdar@physik.uni-halle.de}

\begin{abstract}
 We predict a fast domain wall (DW)
  motion induced by a thermal gradient across a nanoscopic
  ferromagnetic stripe of MnBi. The driving mechanism is an exchange
  torque fueled by magnon accumulation at the DWs. Depending on the
  thickness of the sample, both hot-to-cold and cold-to-hot DW motion
  directions are possible. The finding unveils an energy efficient way
  to manipulate DWs as an essential element in magnetic information
  processing such as racetrack memory.
\end{abstract}
\begin{document}

\flushbottom
\maketitle

\thispagestyle{empty}

\noindent

\section*{Introduction}

Domain walls (DWs), i.e. the cross border of regions with homogeneous but differently
oriented magnetization play a central role in magnetism
\cite{HuSch98}. A particularly interesting suggestion is to exploit
DWs for high density storage in a "racetrack" shift memory
\cite{HaTh08,PaHa08}. The shifting is brought about by passing a
spin-polarized current that exerts a torque on the DWs
\cite{Berg96,Slon96}. While the proposal is technologically attractive
it is hampered by large energy dissipation due to the high current
densities needed. Recent advances resulted in an increase of the DW's
velocity at lower current densities \cite{PhPu15,YaRy15,PaYa15}.
It is however of interest to explore qualitatively new ways for
controlling DWs. Here we show that magnonic current may serve as an
efficient tool for the driving of DWs.

When a temperature gradient is applied to the system, it generates a
magnon current and acts with a torque on the DWs \footnote{No voltage bias is applied. The drift motion of the charge carriers does not affect the magnetic configuration. We note, the DWs considered here are macroscopically large on the scale of the Fermi wave length~\cite{Tatara2}.}. Strong magnetocrystalline anisotropy
impedes motion of DWs. Thus, the choice of material is
essential. So far, only materials with a weak magnetocrystalline
anisotropy received more attention. A particular example is magnetic
nanowires of permalloy
($\texttt{Ni}_{\texttt{0.8}}\texttt{Fe}_{\texttt{0.2}}$)
\cite{ThHa06}. Yttrium iron garnet (YIG)
$\texttt{Y}_{\texttt{3}}\texttt{Fe}_{\texttt{2}}\big(\texttt{FeO}_{\texttt{4}}\big)_{\texttt{3}}$,
$\texttt{Y}_{\texttt{3}}\texttt{Fe}_{\texttt{5}}\texttt{O}_{\texttt{12}}$
is frequently used for thermal activation of spin currents (called
spin Seebeck effect SSE) \cite{KiUch12,UchTa08,JaYa10,UchXi10} and has
a slightly higher anisotropy ($K_{\mathrm{u} 1}(\mathrm{YIG})=-2.0\cdot 10^3$~J/m$^3$ vs.
$K_{\mathrm{u} 1}(\mathrm{permalloy})=-1.0\cdot 10^3$~J/m$^3$)
\cite{Coey10}.

Here we focus on manganese-bismuth compound $\texttt{MnBi}$, generally
known as a hard ferromagnet \cite{Coey14}, where due to a strong
magnetocrystalline anisotropy the ground state possesses an
out-of-plane magnetization orientation. In addition, a remarkable
behavior for the temperature dependence of the magnetic anisotropy of
$\texttt{MnBi}$ was observed: Experiments and first-principles
calculations reveal that the out-of-plane magnetic anisotropy in
$\texttt{MnBi}$ increases at elevated temperatures
\cite{Stutius1974,AnAn14}. As shown here, the notable temperature
dependence of the anisotropy leads to a novel physical phenomena such
as the acceleration of the domain wall motion.  Increasing the applied
temperature gradient in $\texttt{MnBi}$ results in two distinct
effects: It increases the magnonic spin current that drives the domain
wall and creates gradient of the magnetocrystalline anisotropy. The
width of a DW scales according to
$\delta \sim \sqrt{A/K_{\mathrm{u} 1}}$. Here $A$ is the exchange
stiffness. A large magnetocrystalline anisotropy results in relatively
narrow domain walls, i.e. a sharper noncollinear magnetic order and a
larger exchange energy between the neighboring magnetic moments
$-A\vec{M_{n}}\vec{M_{n+1}}$. Hence, the energy landscape forces the
motion of the DW to the area of a lower magnetocrystalline anisotropy
and a lower exchange energy. \chg{Thus, the temperature dependence of
  the magnetic anisotropy in the hard ferromagnet $\texttt{MnBi}$
  generates a fundamentally new type of spin torque, which was not
  studied before. The magnetocrystalline anisotropy torque acts in
  addition to the applied thermal bias leading to a substantial
  enhancement of the DW's velocity.}

\begin{figure}[htb]
    \centering
    \includegraphics[width=0.7\linewidth]{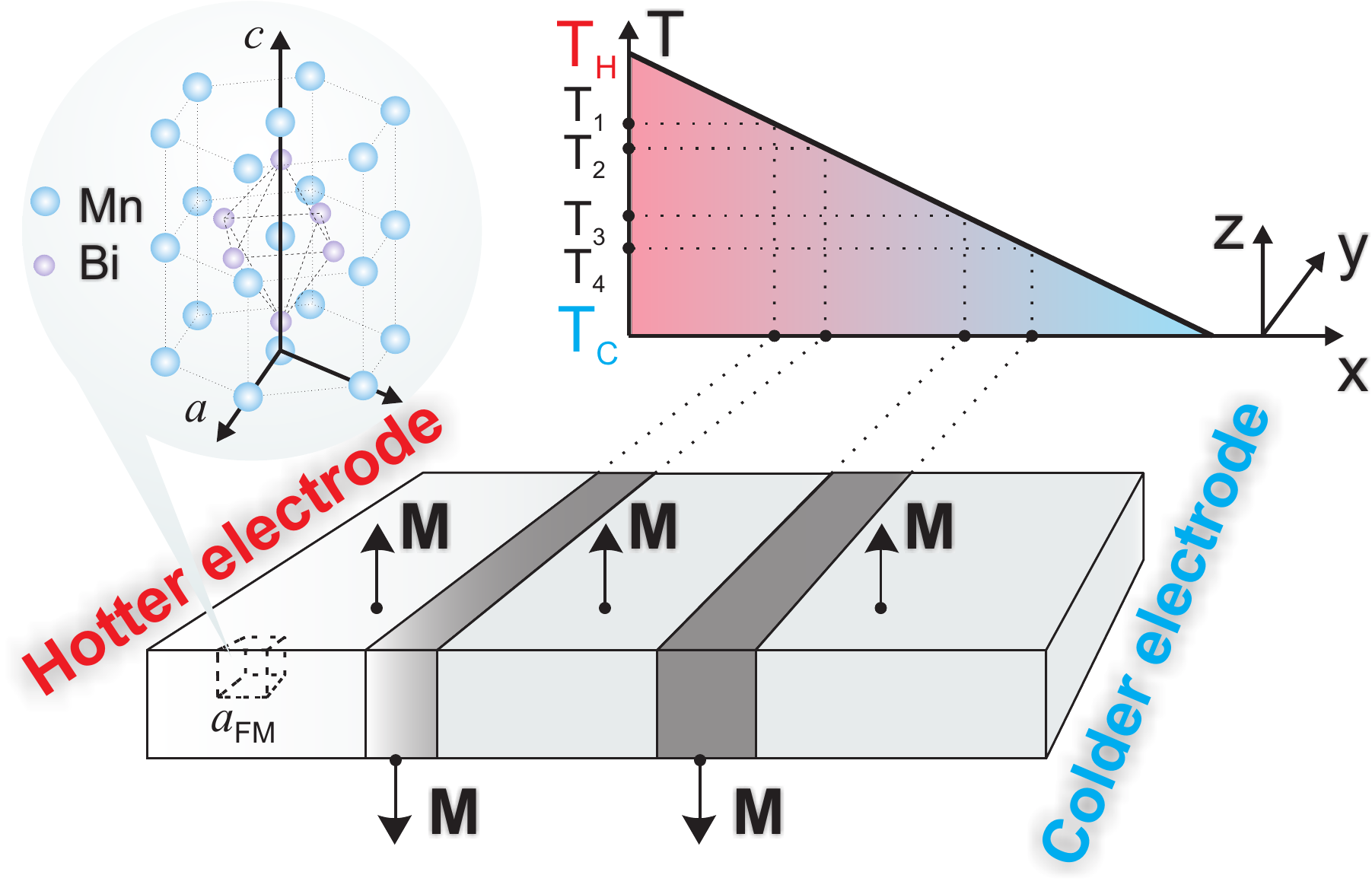}
    \caption{\label{fig1} Schematics of the considered structure with
      a thermal bias applied along the x-axis. $\vec{M}$ indicates the
      direction of the magnetization within a domain,
      $a_{\mathrm{FM}}=1$~nm represents the coarse-grained FM cell
      including the average over several MnBi unit cells, whose
      crystallographic $c$-axis coincides with the z-axis.}
  \end{figure}

\chg{For a comprehensible study of the thermally activated DW motion
  all possible effects related to the applied thermal gradient
  should be addressed. An applied thermal bias generates a magnonic
  spin pumping current (a flux of magnons directed from the hot to the
  cold edge of the sample). On the path to the cold edge the magnonic
  current traverses  the DW. Our calculations (see bellow) show that
  under certain conditions  two different scenarios of the DW motion
  are realized: If the width of the DW is  small enough (less than
  7nm, see Fig.2, Fig.3 in the supplementary information) the DW is
  transparent for thermal magnons and the magnons pass through the DW
  without a sizeable change of magnons's momentum. Naturally, the spin
  pumping current does not exert a magnonic pressure on the DW's
  surface, while  the angular momentum is still transferred (the
  angular momentum is changed by $2\hbar$ and the momentum of the
  magnon is conserved). A shortage of the angular momentum appears in
  the vicinity of the hot edge, while an extra angular momentum is
  accumulated at the cold edge of the sample (magnon accumulation
  effect, see \cite{Ritzmann2}). In order to compensate for the
  imbalance in the distribution of the angular momentum, the DW
  propagates in the direction opposite to that of the magnon
  \cite{YaWa11,TaKo04}. As for the direction of the thermally-induced
  DW's motion, one should also consider the entropic torque. The free
  energy  $F = \bar E - TS$ (here $\bar E$ is the internal energy and
  $S$ is the entropy) is minimized at elevated temperatures $T$. Thus,
  the entropic torque also provokes a motion of DW towards the hotter
  edge \cite{HiNo11,SchRi14,WaWa14,KoTs12,JiUp13}.  One may further
  argue that the DW is not thermally isolated and the heat flux
  associated  with  the magnonic current influences the entropy of the
  DW \cite{YaCa15}. However, the influence of the entropic torque on
  the DW motion is an established  trend  with the  DW  moving towards
  the hot edge. \\
%
For a DW width exceeding 15nm, the magnonic spin pumping current is
totaly reflected by the DW. Therefore, in this case a second scenario
is relevant: The magnons exert a sustainable pressure on the DW's
surface.} If the magnonic pressure is strong enough,  a thermally
activated spin pumping current drags the DW  to the cold edge. 
\chg{Our full numerical calculations evidence}  that indeed the direction of the
DWs motion is system dependent. The microscopic theory \cite{YaKa13}
predicts a strong magnonic recoil effect for the DW motion.  The momentum
of the DW depends on the tilt angle of its plane. If a magnon is
reflected by the DW, the conservation of the total angular momentum
induces a rotation of the DW. Thus, the magnon reflection leads to a
motion of the DW in the direction of magnonic current
\cite{YaKa13,YaBa12}. The pressure exerted by the magnonic current on
the DW reads: $P=2\delta nv_{k}\hbar k$. Here $v_{k}=\mid
\vec{v_{k}}\mid$, $k=\mid \vec{k}\mid$ are the velocity and the wave
vector of magnons, $\delta n=n_{neq}\big(T\big)-n_{eq}\big(T\big)$
quantifies the magnon accumulation effect, i.e. the excess of the
density of non-equilibrium magnons $n_{neq}\big(T\big)$ compared to
the reference number of equilibrium magnons $n_{eq}\big(T\big)$ at the
same temperature, but in the absence of the thermal gradient. Exerted pressure is proportional
to the applied thermal bias.  Our calculations (see Fig. 1 of Suppl. Information) are in line with the
fact that magnons created in the hot area propagate toward the cold
area. The left (right) side of the DW's (see Fig. \ref{fig1}) is
associated with the high (low) temperature of the applied thermal
gradient. Therefore, the magnon accumulation is positive $\delta n>0$ at
the left side of the DWs and is negative $\delta n<0$ at the right
side exerting a magnonic pressure that drives the DWs. The direction
of the motion of DW  depends considerably on the ratio between the
reflected and the transmitted magnons. The velocity of the DW can then
be evaluated according to \cite{WaGu12}
\begin{equation}
V_{\mathrm{DW}}=-\frac{R}{1+\alpha^{2}}U+\frac{\alpha \big(1-R\big)\Delta k}{1+\alpha^{2}}U.
\label{eq_1}
\end{equation}
Here $U=\big(\gamma \hbar n v_{k}\big)/(\mu_{B}M_{s})$, $\alpha$ is the phenomenological Gilbert damping, $\gamma$ is
the gyromagnetic ratio, $M_{\mathrm{S}}$ is the saturation magnetization, \chg{$\Delta$ is the width of the DW and $k$ is the
magnon's wave vector}, and $0<R<1$ is
the transmission coefficient. Driving DW via the spin waves excited by
a microwave field shows a strong frequency dependence \cite{YaWa11}. \chg{In contrast to
the field-driven DW motion \cite{Bauer2}, the thermally activated magnonic spin current is non-monochromatic\cite{Chotorlishvili2}.}
Spectral characteristics of the magnons contributing to the magnonic
accumulation are not yet fully settled. The magnonic current could be
due to the difference between the magnon and the phonon temperatures
\cite{XiBa10}. Recent spatially resolved experiments carried out via
Brillouin light scattering technique \cite{AgVa13} show a
non-vanishing spin current for equal magnon and phonon temperatures.
However, as shown in \cite{EtCh14}, the thermally activated spin
current is dictated by the magnon temperature profile rather than by
the difference between the phonon and the magnon temperatures. In a
recent experiment \cite{BoHe14} low frequency subthermal magnons were
identified as a source for the spin current. These are magnons with a
frequency lower than the internal cutoff frequency
$\omega=\gamma\left(H_{\mathrm{eff}}+\frac{2A}{M_{\mathrm{S}}}k^2\right)<\omega_{\mathrm{c}}$
(here $A$ is the exchange stiffness and $H_{\mathrm{eff}}$ stands
for the effective magnetic field).
\chg{The internal cutoff frequency is not related to the structure of the DW but to the spectrum of the thermally excited magnons. The frequency spectrum of the thermally excited magnons is not monochromatic and covers a certain frequency domain. The internal cutoff frequency $\omega_{c}$  is the highest frequency of the magnons, which are excited thermally. Hence, the magnons with frequencies higher than the internal cutoff frequency $\omega > \omega_{c}$ do not contribute to the thermally activated magnonic current \cite{BoHe14,Chotorlishvili5}. We observed that magnons can either pass through the DW or are reflected off the  DW depending on the geometry (thickness) of DW itself, while the spectral characteristics of the magnons (defined by thermal bias) are irrelevant.} We find that in the case of a non-monochromatic, thermally  activated magnonic current the ratio
between the reflected and the transmitted magnons  depends crucially on the characteristics of the DW itself, rather than on the cutoff frequency
$\omega_{\mathrm{c}}$. For the same magnonic current we detect DWs with different geometry moving in the opposite direction.
\section*{Results}
\begin{center}
  \begin{figure*}[htb]
   \centering
   \includegraphics[width=.85\textwidth]{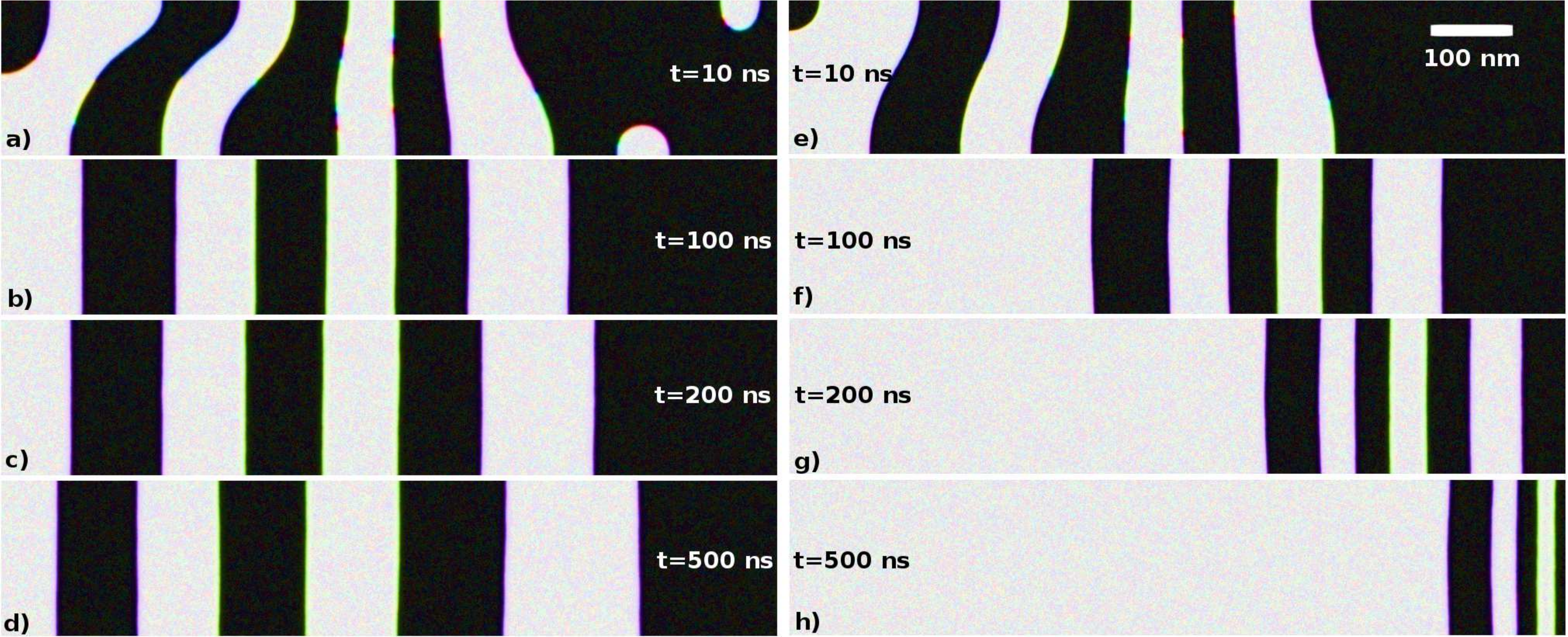}
   \caption{\label{fig2} Top view of the magnetization configuration
     with no temperature gradient applied $\nabla _{\mathrm{x}} T=0$
     (left panel, a-d) and the magnetization configuration for both
     $\nabla _{\mathrm{x}} T=0.15$~K/nm and
     $K_{\mathrm{u} 1}(T)=(9.69 \cdot 10^3 T - 1.12 \cdot 10^6)$
     J/m$^3$(right panel, e-h) of a $1000$nm$\times
     200$nm$\times 25$nm
     MnBi sample at $T_{C}=300$~K and different time moments in the range
     [$0:500$]~ns. The light color represents the magnetization pointed
     towards the reader and the dark color indicates the magnetization
     away from the reader both perpendicularly to the surface of the
     figure (cf. Fig. \ref{fig1}). The magnetization configuration is
     calculated including long-range interactions (the demagnetizing
     fields). The anisotropy strength is set according to
     first-principles calculations of Ref. \cite{AnAn14} fitted for
     the interval of [$300:400$]~K (right panel, e-h).
    \chg{A thermal bias is applied to the sample such that $T_{H}-T_{C}=150$~K. All simulations were performed  with the initial state at $t=0$ being always chosen randomly with regard to the magnetization orientation. Then, the magnetization of a given sample was relaxed up to 10ns in the absence of an external thermal bias.}}
   \end{figure*}
\end{center}

\begin{center}
  \begin{figure}[htb]
   \centering
   \includegraphics[width=.7\textwidth]{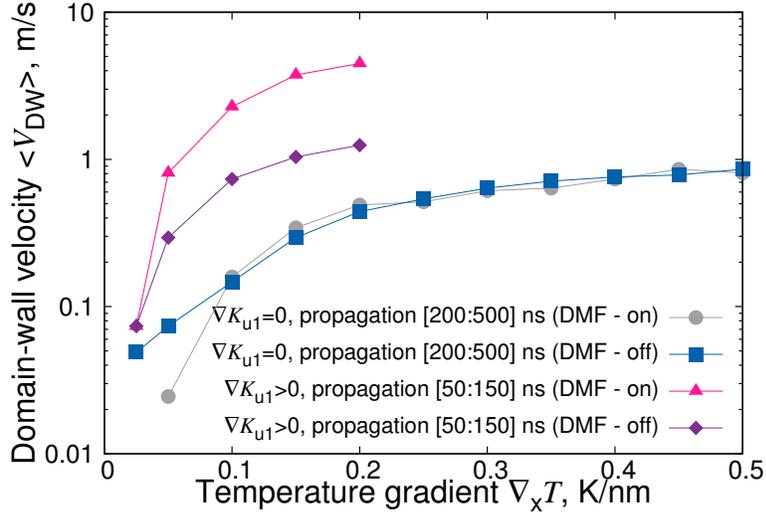}
   \caption{\label{fig3}
    Dependence of the average domain-wall velocity on the
     temperature gradient $\nabla _{\mathrm{x}} T$ for a
     $1000$nm$\times 200$nm$\times 25$nm MnBi sample. The parameters
     are taken from Table 1 (Supplem. Information) and for $\alpha=1.0$. "DMF" stands for
     demagnetizing fields which are taken (or not) into account for the
     respective curves. Solid circles and squares represent the data
     for zero anisotropy gradient $\nabla K$. Triangles and diamonds
     show the situation with non-zero anisotropy gradient
     according to the first-principles calculations of Ref.
     \cite{AnAn14} fitted for the interval of [$300:400$]~K as
     $K_{\mathrm{u} 1}(T)=\big(9.69 \cdot 10^3 T - 1.12 \cdot 10^6$ \big) J/m$^3$.
\chg{ The increase of the DW's velocity is a cooperative effect of the anisotropy gradient and the demagnetization fields. When the magnetocrystalline anisotropy gradient is applied the demagnetization field contributes to the formation of domains. We observed (not shown) that domains in this case are smaller and this enhances their mobility.}}
   \end{figure}
\end{center}

As shown in Figs. \ref{fig2}, a)-d), a
homogeneously heated sample does not lead to a directed domain wall
motion in the sample, but rather a redistribution of domains on a long
time scale is observed. Considering the time laps of $[10:500]$ ns
while a linear temperature gradient is applied and accounting also for
the temperature dependence for the magnetocrystalline anisotropy
\cite{AnAn14} (the anisotropy is stronger for higher temperatures), we
observe (cf. Fig. \ref{fig2}, e)-h)) that almost all domains reach
the right edge of the sample. Also the size of FM domains is reduced
during their motion.  The pressure exerted by the magnon accumulation effect modifies the shape of DWs.
Fig.~\ref{fig3} illustrates the domain wall
motion. The averaged velocity of the domain wall motion
$<V_{\mathrm{DW}}>$ is plotted as a function of the linear temperature
gradient. As expected, the velocity grows with elevating the
temperature gradient, however it shows a slight saturation at higher
gradients. The reason for that is the saturation of the magnonic
current \chg{which was theoretically } predicted in Ref.~\cite{Chotorlishvili}.
Surprisingly, the demagnetizing fields do not influence much the
domain wall speed for a homogeneous anisotropy (Fig.~\ref{fig3},
circles and squares). The inhomogeneity of the anisotropy induced by the
temperature gradient increases the speed of domain wall motion by
approximately one order of magnitude.

\begin{center}
  \begin{figure}[htb]
   \centering
   \includegraphics[width=.7\textwidth]{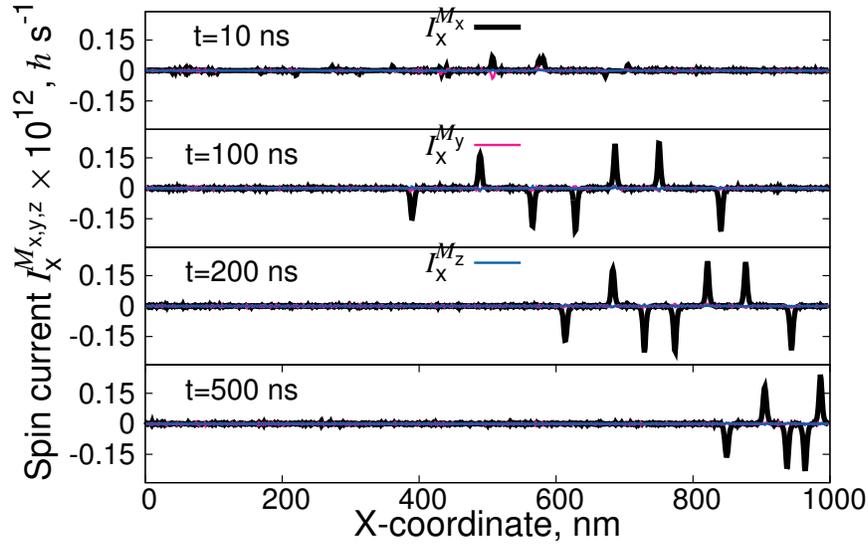}
   \caption{\label{fig4} { Time-resolved magnon spin currents
     corresponding to magnetic configurations shown in Fig.
     \ref{fig2} (right panel) of a $1000$nm$\times 200$nm$\times 25$nm MnBi sample
     for the temperature gradient of $0.15$ K/nm at different times in
     the range [$0:500$]~ns. The left side of the sample is hotter,
     the right edge is always kept at room temperature. The
     magnetization configuration is calculated for \textit{non-zero}
     demagnetizing fields. Anisotropy gradient is chosen according to
     first-principles calculations of Ref. \cite{AnAn14} fitted
     for the interval of [$300:400$]~K in the form
     $K_{\mathrm{u} 1}(T)=\big(9.69 \cdot 10^3 T - 1.12 \cdot 10^6$ \big)J/m$^3$.}}
   \end{figure}
\end{center}

\begin{center}
  \begin{figure}[htb]
   \centering
   \includegraphics[width=.7\textwidth]{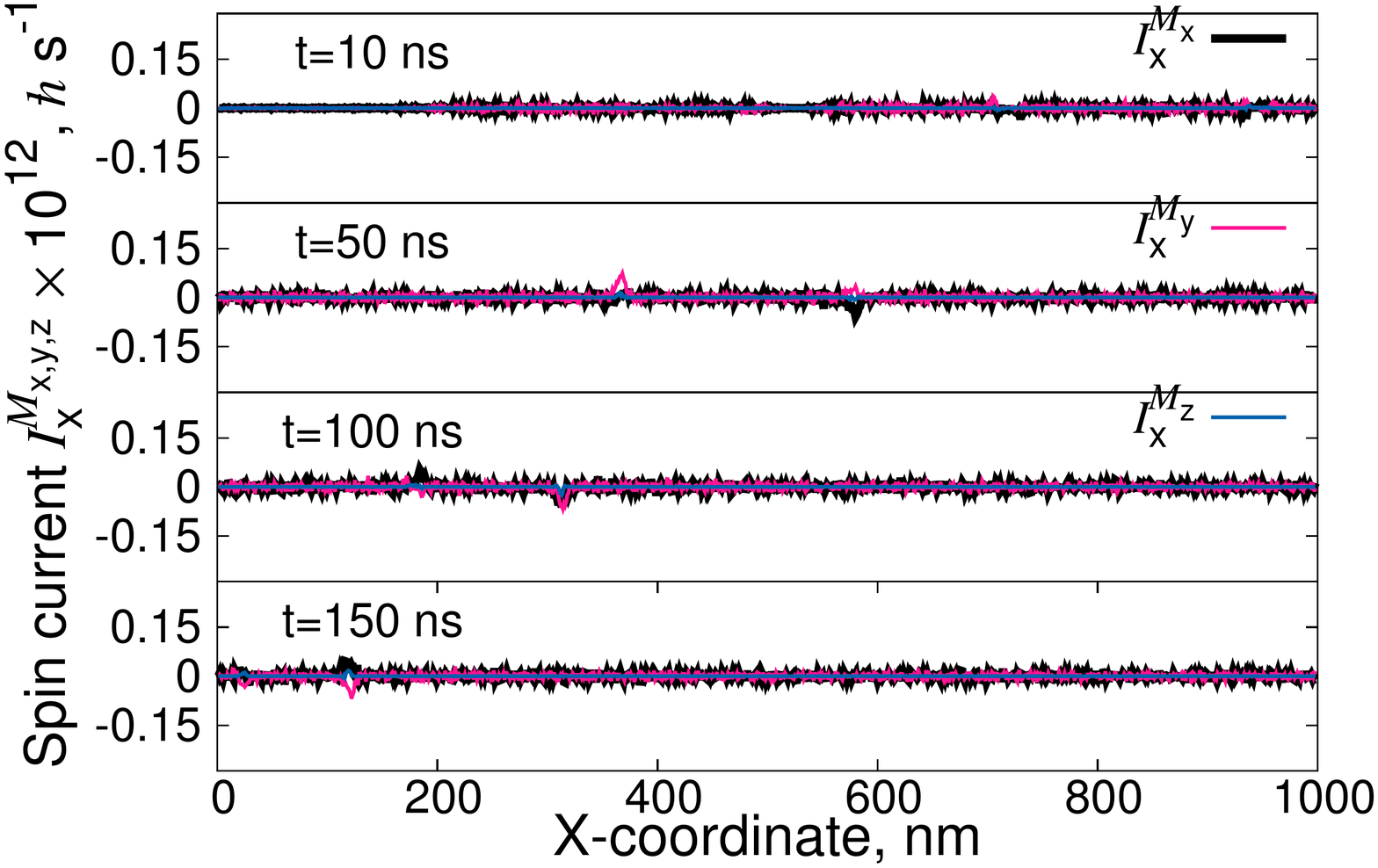}
   \caption{\label{fig5} {Time-resolved magnon spin currents of a
     $1000$nm$\times 200$nm$\times 1$nm MnBi sample for the
     temperature gradient of $0.15$ K/nm at different times in the
     range [$0:150$]~ns. The left side of the sample is hotter, the
     right edge is always kept at room temperature. The magnetization
     configuration is calculated for \textit{non-zero} demagnetizing
     fields. The anisotropy is homogeneous, i.e. $K_{\mathrm{u} 1}=1.75\cdot 10^6$~J/m$^3$}.}
   \end{figure}
\end{center}

Our general interpretation of
the obtained DW dynamics relies on the definition of the magnon spin
current given in Ref.~\cite{EtCh14} by a recursive formula\footnote{Recurcive relation for the spin current:
 $\displaystyle I^{\alpha}_{n}=I^{\alpha}_{0}-\frac{2Aa}{M_{S}^{2}}\sum_{m=1}^{n}M^{\mathrm{\beta}}_{\mathrm{m}}(M^{\mathrm{\gamma}}_{\mathrm{m-1}}+M^{\mathrm{\gamma}}_{\mathrm{m+1}})\varepsilon_{\alpha\beta\gamma},
$ where $\varepsilon_{\alpha\beta\gamma}$ is the Levi-Civita antisymmetric tensor, $A$ effective exchange stifness, $a$ lattice constant, and $M_{S}$ saturation magnetization.  Greek indexes define the current components and the Latin ones denote sites.}. The
x-component of the spin current
$I^{M_{\mathrm{x, y, z}}}_{\mathrm{x}}(t)$ is calculated in a
time-resolved manner. Fig.~\ref{fig4}  demonstrates clearly that most
crucial for the magnon spin current is the presence of FM domain
walls, where different components of the magnon-current tensor show
pronounced peaks. The accumulation of the magnon spin current at the
domain walls leads to a torque acting on the domain walls, shifting
them towards the colder edge. The situation might be different if the
thickness of the sample is drastically reduced. In this case the
\chg{magnonic} spin current penetrates  the DWs (Fig. \ref{fig5}) and
the entire domain walls move towards the hotter electrode. However in
this case the DWs move opposite to the anisotropy gradient. Therefore,
the velocity of the DWs decreases. 

\section*{Discussion}

We studied thermally activated motion of the DWs in the
manganese-bismuth compound MnBi, generally known as a hard
ferromagnet.  We found that the unusual dependence of the
magnetocrystalline anisotropy on the temperature (magnetocrystalline
anisotropy in MnBi increases at elevated temperatures) is advantageous
for a fast thermal steering of DWs. Sharp noncollinear magnetic orders
formed under the strong magnetocrystalline anisotropy are
energetically unfavorable. Therefore, the DWs slip to areas of small
magnetocrystalline anisotropy.  Thus, the gradient of the
magnetocrystalline anisotropy acts as a macroscopic driving assisting
the thermally activated motion of the DWs.  The microscopic mechanism
for the DWs motion is based on the magnonic spin current. The
reflected magnonic spin current drags the DW to the cold edge while
transmitted current forces the DW to the hot edge. We also observe a
deformation of the shape of DWs due to the pressure exerted by the
accumulation of magnons.

\section*{Methods}

To provide parameters for our model simulations we performed extensive
first-principles study of MnBi using a self-consistent Green function
method within the framework of the density functional
theory~\cite{Lueders2001}. The ground state properties were calculated
for the experimental lattice constant, while for the magnetic
anisotropy energy (MAE) calculations the crystalline structure
parameters were adopted from experiments, which studied the dependence
of the MAE on the applied temperature \cite{Stutius1974}. First of
all, we found that our simulations within the generalized gradient
approximation provide an adequate description of MnBi. The main ground
state properties, obtained within our calculations, are listed in the
Table 1, second column (Supplem. Information). Thus, in accordance with
experiments, MnBi is a robust ferromagnet with Curie temperature of
$680$~K. The magnetic moments and exchange interaction parameters,
estimated within the magnetic force theorem, provide the exchange
stiffness close to the experimental values (cf. Table 1 (Supplem.
Information)). The MAE calculated for $T=0$~K and for higher
temperatures is in a good agreement with the experimental results and
other theoretical studies \cite{Stutius1974,AnAn14}. The influence of the conductance electrons
is incorporated in the effective exchange stiffness \cite{Liechtenstein}.

With the microscopic parameters delivered by experiments and first
principles calculations we performed micromagnetic simulation for the
magnetization dynamics on the basis of Landau-Lifshitz-Gilbert (LLG)
equation, as implemented in the \texttt{mumax3}-micromagnetic
simulation package \cite{VaLe14}.

\section*{Supplementary Information}

Table \ref{table_1} gives an overview over the main parameters that are relevant for the magnetic MnBi-alloy. In the present calculations the parameters termed as "experimental values" (third column) were employed. One of the main parameter that is important for the magnetic configuration is the ferromagnetic (FM) cell $a_{\mathrm{FM}}$. It should be chosen such that detailed magnetic structures such as FM domains can be resolved. It is known \cite{Coey10} that the width of FM domain walls scales according to $\delta \sim \sqrt{A/K_{\mathrm{u} 1}}$, where $A$ is the exchange stiffness.
 Since  the magneto crystalline anisotropy $K_{\mathrm{u} 1}$
  in MnBi is very high, it results in a relatively narrow domain walls with widths around $2  - 6$ nm. To resolve  fine structures within the domain walls we deemed, upon tests, that $a_{\mathrm{FM}}=1$ nm as appropriate. \textcolor{black}{In the \emph{ab-initio}-calculations we obtained the value for the saturation magnetization\footnote{Calculated as $M_{\mathrm{S}}=\frac{\mu_{S}}{V_0}$, where $V_0=137.48$ \AA$^3$ \,\, is a volume of the unit cell and $\mu_{\mathrm{S}}(T=0 \mathrm{K})=8.4 \mu_{B}$ or $\mu_{\mathrm{S}}(T=300 \mathrm{K})=7.33 \mu_{B}$ for MnBi.} and the exchange stiffness constant\footnote{Calculated based on the ratio $A=\frac{DM_{\mathrm{S}}}{2g\mu_{\mathrm{B}}}$ (in MnBi $D=670$ meV \AA$^2$), which is derived in Ref. \cite{HaGa09}.} shown in Table \ref{table_1}. For comparison, analogous values for the exchange stiffness\footnote{Calculated from the ratio for the FM domain wall thickness $\delta_{\mathrm{FM}}=\sqrt{A/K_{\mathrm{u} 1}}$ based on values from Ref. \cite{GuCh92} at room temperature.}, the Gilbert damping\footnote{Calculated based on the ratio $\alpha=\frac{\sqrt{3}}{2}\frac{\gamma}{\omega} \Delta B_{\mathrm{pp}}$, which is approximated for the half-height of the width of the FMR spectrum curve at $T=130$~K. The values of $\Delta B_{\mathrm{pp}}\approx 0.05$~T, $\omega/(2\pi)=9$~GHz are taken from Ref. \cite{ChGo64}. Gyromagnetic ratio is $\gamma=1.76\cdot 10^{11}$~(Ts)$^{-1}$.} and the FM cell size\footnote{Should be of the order of $2.5$ nm, which is min. $\delta_{\mathrm{FM}}$ (T=300 K) based on values of Ref. \cite{GuCh92}.} are also listed in Table \ref{table_1}.}

The size of the simulated MnBi sample has the dimension of $1000$ nm along the x-axis, $200$ nm along the y-axis, and  a thickness of $25$ nm (Sections I, II and III). It required  $1.2$ GB of the GPU-memory  while including  demagnetizing fields. To examine the dynamics of domain walls we focused  on an stripe sample with the temperature gradient being along the stripe (chosen as the x-direction). The value $25\div 30$ nm for the sample thickness was taken in accordance with the minimal known thickness where a magnetic pattern was still recognizable \cite{Dekk74}.

In order to initiate FM domains, we proceed in a natural way by  starting from a random magnetic configuration at zero magnetic field and at room temperature ($T=300$ K). To achieve convergence to a well-defined magnetic configuration a relatively large damping was chosen
 which resulted in a convergence to the ground state on the time scale of several to ten nanoseconds
 (simulation time step was set to $5\cdot 10^{-13}$ s). After reaching the ground state which is a magnetic pattern with several FM domain walls (typically after $10$ ns), the temperature gradient is applied, where the hotter end is the left part of the sample, whereas the right end is always kept at room temperature.

The magnetic configurations with time are obtained from the solution of the Landau-Lifshitz-Gilbert (LLG) equation within the micromagnetic framework by means of \texttt{mumax3}-software \cite{VaLe14}.

\begin{center}
 \begin{table*}[htb]
 \begin{tabular}{lll} 
  \hline
  \hline
 DESCRIPTION & \hspace{2.ex} \textit{ab-initio} \hspace{2.ex} & \hspace{2.ex} EXPERIMENT \hspace{2.ex} \\
  \hline
  Magnetization $M_{\mathrm{S}}$, [A/m] \tabrule& $5.6\cdot 10^5$ ($T=0$~K) & $7.42\cdot 10^5$ (T=0 K) \cite{GuCh92}  \\
  \tabrule & $4.9\cdot 10^5$ ($T=300$~K) & $6.40\cdot 10^5$ (T=300 K) \cite{GuCh92} \\
  Exch. stiffness $A$, [J/m] \tabrule& $1.43\cdot 10^{-11}$ & $6.25\cdot 10^{-12}$ \\
  Anis. strength $K_{\mathrm{u} 1}(T)$, [J/m$^3$] \tabrule& - $0.44\cdot 10^6 (T=0 \, \mathrm{K})$ \cite{Antropov} & \\
    \tabrule & $1.4\cdot 10^6 (T=300 \, \mathrm{K})$ \cite{Antropov} & $1.75\cdot 10^6 (T=300 \, \mathrm{K})$ \cite{GuCh92} \\
  Anisotropy type \tabrule& uniaxial & uniaxial\\
  Gilbert damping $\alpha$  \tabrule& - & 0.0054 (T=130 K)\\
  Curie-Temperature $T_{\mathrm{C}}$, [K] \tabrule& $680$ & $775$ (Ref. \cite{GuCh92})\\
  FM cell size $a_{\mathrm{FM}}$, [nm] \tabrule & - & $2.5$\\
  \hline
  \hline
 \end{tabular}
 \caption{Parameters relevant for MnBi-alloy.}
 \label{table_1}
\end{table*}
\end{center}

\subsection*{Domain-wall motion}
\begin{center}
  \begin{figure}[htb]
   \centering
   \includegraphics[width=.7\textwidth]{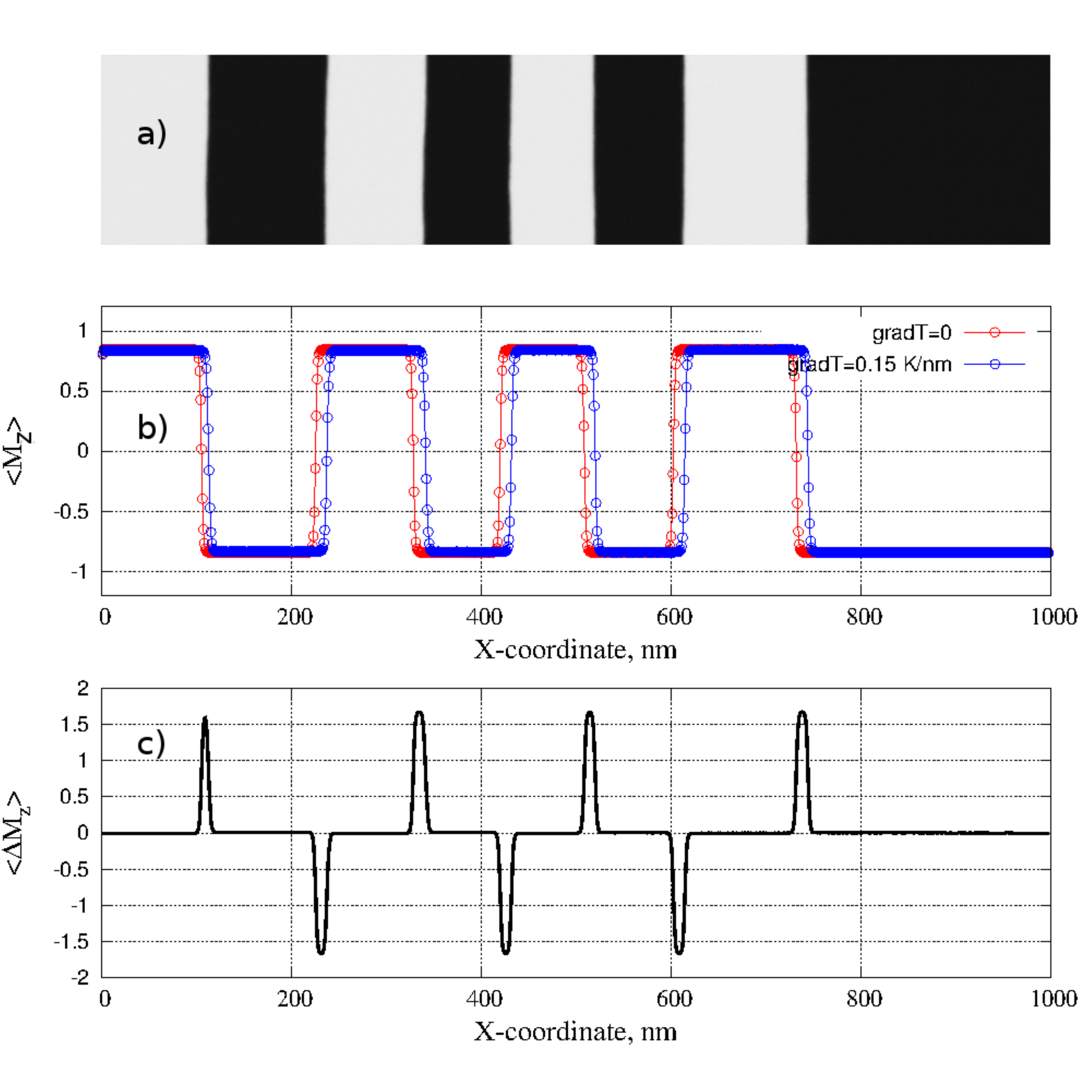}
   \caption{\label{fig_2} (a) Top view of the magnetization configuration (black color indicates the magnetization direction "down", white color implies the magnetization direction "up") of a $1000$nm$\times 200$nm$\times 25$nm MnBi sample for the temperature gradient of $0.15$ K/nm at time $100$~ns. The spin accumulation (c) is calculated according to the expression $<\Delta M^n_{\mathrm{z}}> = < M^n_{\mathrm{z}}>|_{\bigtriangledown T \neq 0} - < M^n_{\mathrm{z}}>|_{\bigtriangledown T = 0}$ for a $1000$nm$\times 200$nm$\times 25$nm MnBi sample, where $< M^n_{\mathrm{z}}>|_{\bigtriangledown T = 0}$ (b) stands for the data with no temperature gradient at time $t=100$~ns and $< M^n_{\mathrm{z}}>|_{\bigtriangledown T \neq 0}$ (b) is for the data with a temperature gradient $0.15$~K/nm  at time $t=100$~ns. (c) A positive spin accumulation corresponds always to  the left side of domains, negative values of the spin accumulation are associated with the right side of the corresponding domains.}
  \label{fig:dw} \end{figure}
\end{center}

Fig. \ref{fig:dw} provides additional illustration of the DW motion under thermal bias.

\subsection*{Domain-wall thickness}
\begin{center}
  \begin{figure}[htb]
   \centering
   \includegraphics[width=.7\textwidth]{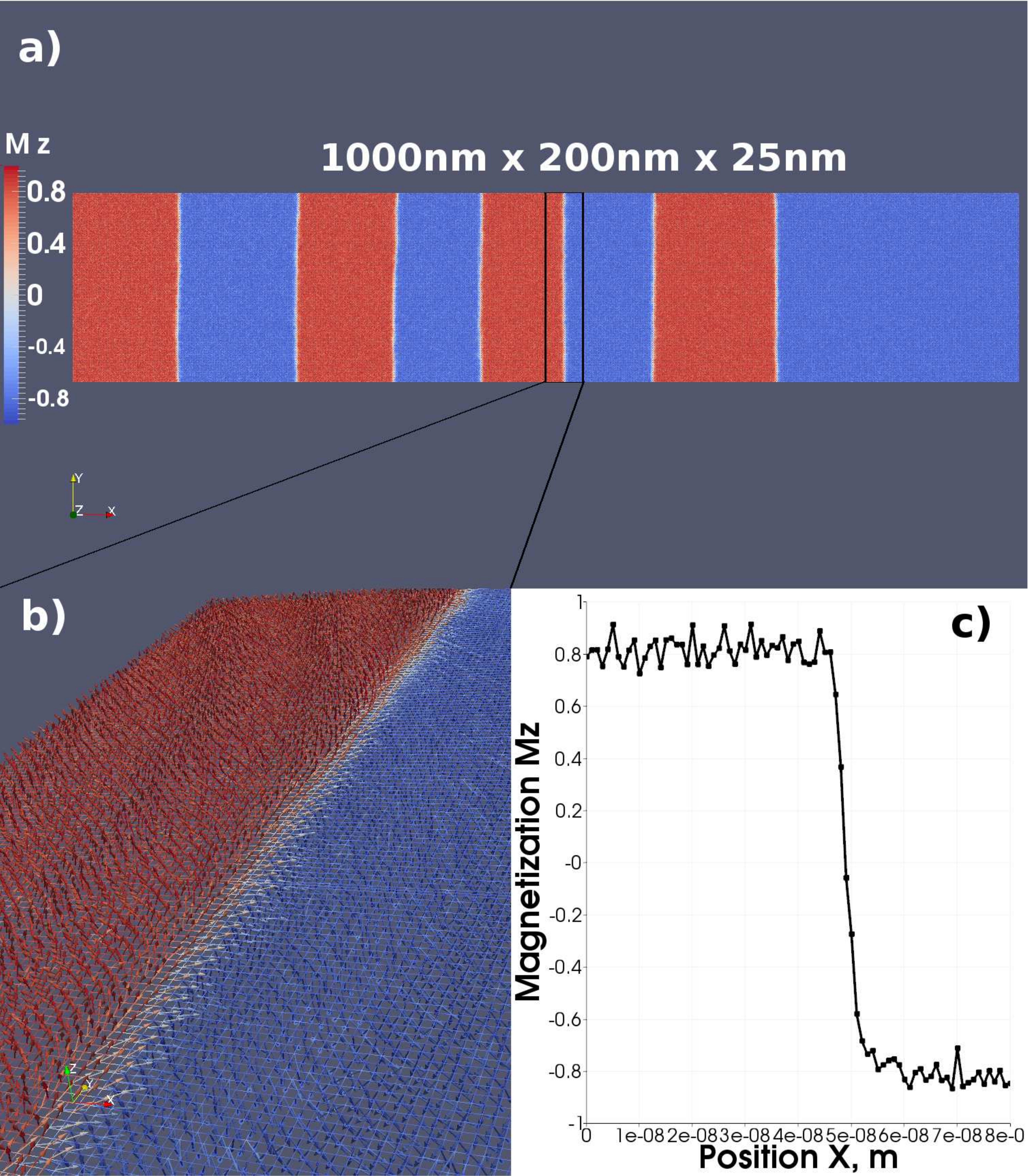}
   \caption{\label{fig_3} (a) Top view of the magnetization configuration (blue color indicates the magnetization direction "down", red color implies the magnetization direction "up") of a $1000$nm$\times 200$nm$\times 25$nm MnBi sample for the temperature gradient of $0.15$ K/nm at time $100$~ns ($\nabla K_{\mathrm{u} 1}=0$, DMF - on). (b) stands for the  magnetization configuration frame between $[470:570]$ nm (top layer only). (c) indicates the averaged magnetization profile along a line drawn parallel to the x-direction and connecting edges of the frame $[470:570]$ nm for the top layer. The resulting domain-wall thickness is $d_{\mathrm{DW}}\approx 15$ nm.}
   \label{fig:dwthick}\end{figure}
\end{center}

\begin{center}
  \begin{figure}[htb]
   \centering
   \includegraphics[width=.7\textwidth]{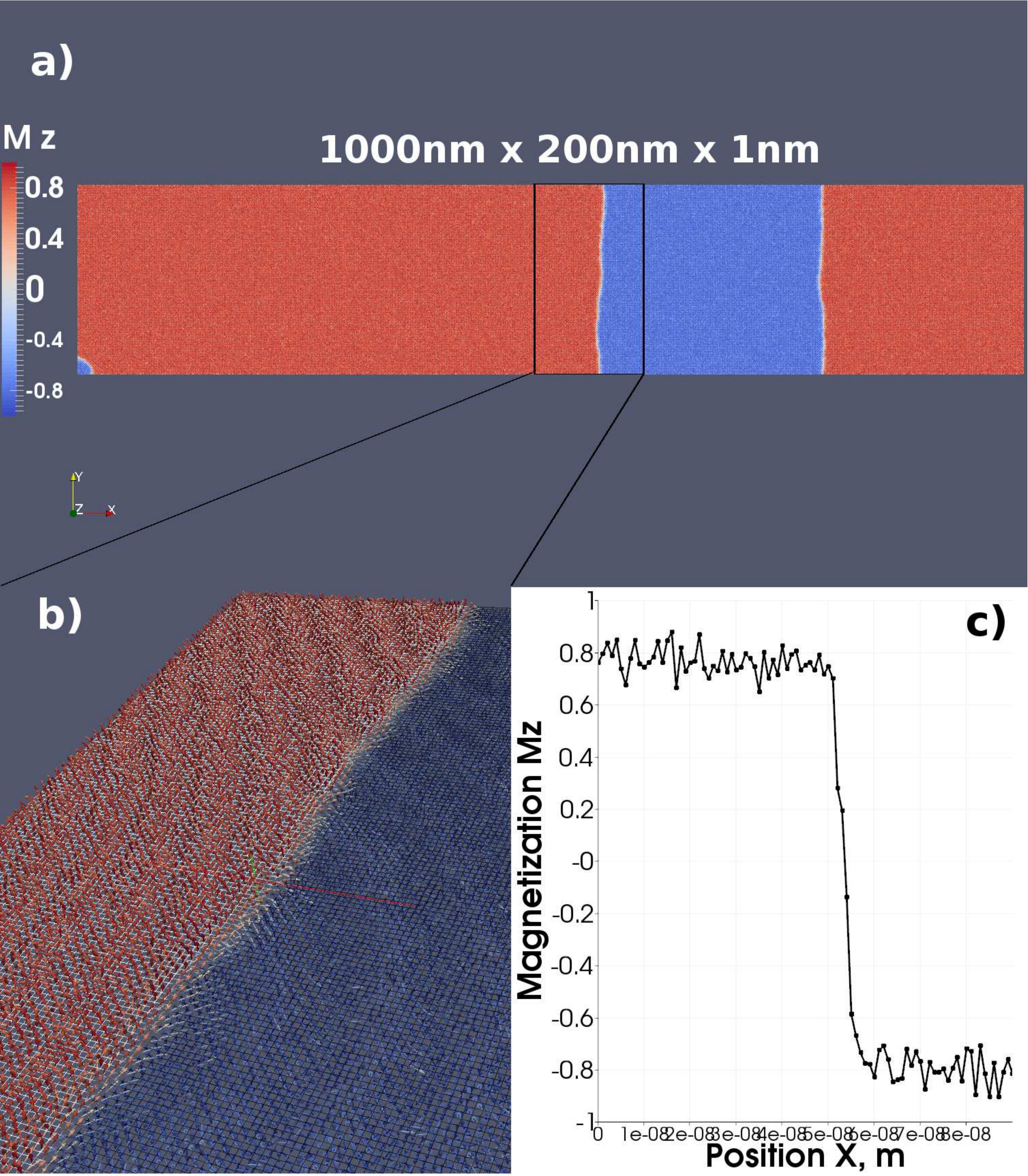}
   \caption{\label{fig_4} Same as in Fig. \ref{fig:dwthick} but (a)  for the temperature gradient of $0.15$ K/nm at time $20$~ns ($\nabla K_{\mathrm{u} 1}=0$, DMF - on). (b)  the  magnetization configuration frame is between $[500:600]$ nm. (c)  the averaged magnetization profile for the frame $[500:600]$ nm. The resulting domain-wall thickness is $d_{\mathrm{DW}}\approx 7$ nm.}
   \label{fig:dwthick2}\end{figure}
\end{center}

In  Figs. \ref{fig:dwthick}, \ref{fig:dwthick2} we demonstrate the trend in the thermally assisted DW motion with the DW thickness.

\section*{Acknowledgements}

This work is supported by the Deutsche Forschungsgemeinschaft under
grants BE 2161/5-1, ER 340/5-2, ME 1153/15-2 and by the Joint
Initiative for Research and Innovation within the Fraunhofer and Max
Planck cooperation program.

\section*{Author contributions statement}

All authors contributed to the discussion and analysis of the research. A. S. performed micromagnetic simulations, L. C. and J. B. provided the
theoretical explanation and compiled the text. A. E, X. Z., S. O., I. M., E. K. U. G. performed DFT calculations and provided parameters for the model.

\end{document}